\documentclass[runningheads]{llncs}
\usepackage[T1]{fontenc}
\usepackage{cite}
\usepackage{graphicx}
\usepackage{algorithmic}
\usepackage{graphicx}
\usepackage{graphics}
\usepackage{textcomp}
\usepackage{subfigure}
\usepackage{tabularx}
\usepackage{multirow}
\usepackage{bm}
\usepackage{latexsym}
\usepackage{booktabs}
\usepackage{threeparttable}
\usepackage{epsf}
\usepackage{algorithm}
\usepackage{color,colortbl}

\usepackage{enumerate}
\usepackage{hyperref}
\hypersetup{hidelinks,
	colorlinks=true,
	allcolors=blue,
	pdfstartview=Fit,
	breaklinks=true}

\begin{document}
\title{A Peer-to-peer Federated Continual Learning Network for Improving CT Imaging from Multiple Institutions}
\titlerunning{icP2P-FL with CL for Improving CT Imaging from Multiple Institutions}
\author{Hao Wang\inst{1,2} \and
Ruihong He\inst{1} \and
Xiaoyu Zhang\inst{1,2}\and
Zhaoying Bian\inst{1,2}\and
Dong Zeng\inst{1,2}\and
Jianhua Ma\inst{1,2}
}
\authorrunning{H. Wang et al.}
\institute{School of Biomedical Engineering, Southern Medical University, \\ Guangdong 510515, China \and
Pazhou Lab (Huangpu), Guangdong 510000, China }

\maketitle            
\begin{abstract}
Deep learning techniques have been widely used in computed tomography (CT) but require large data sets to train networks. Moreover, data sharing among multiple institutions is limited due to data privacy constraints, which hinders the development of high-performance DL-based CT imaging models from multi-institutional collaborations. Federated learning (FL) strategy is an alternative way to train the models without centralizing data from multi-institutions. In this work, we propose a novel peer-to-peer federated continual learning strategy to improve low-dose CT imaging performance from multiple institutions. The newly proposed method is called peer-to-peer continual FL with intermediate controllers, i.e., icP2P-FL. Specifically, different from the conventional FL model, the proposed icP2P-FL does not require a central server that coordinates training information for a global model. In the proposed icP2P-FL method, the peer-to-peer federated continual learning is introduced wherein the DL-based model is continually trained one client after another via model transferring and inter-institutional parameter sharing due to the common characteristics of CT data among the clients. Furthermore, an intermediate controller is developed to make the overall training more flexible. Numerous experiments were conducted on the AAPM low-dose CT Grand Challenge dataset and local datasets, and the experimental results showed that the proposed icP2P-FL method outperforms the other comparative methods both qualitatively and quantitatively, and reaches an accuracy similar to a model trained with pooling data from all the institutions.
\end{abstract}
\begin{figure}
\includegraphics[width=1\textwidth,height=0.6\textwidth]{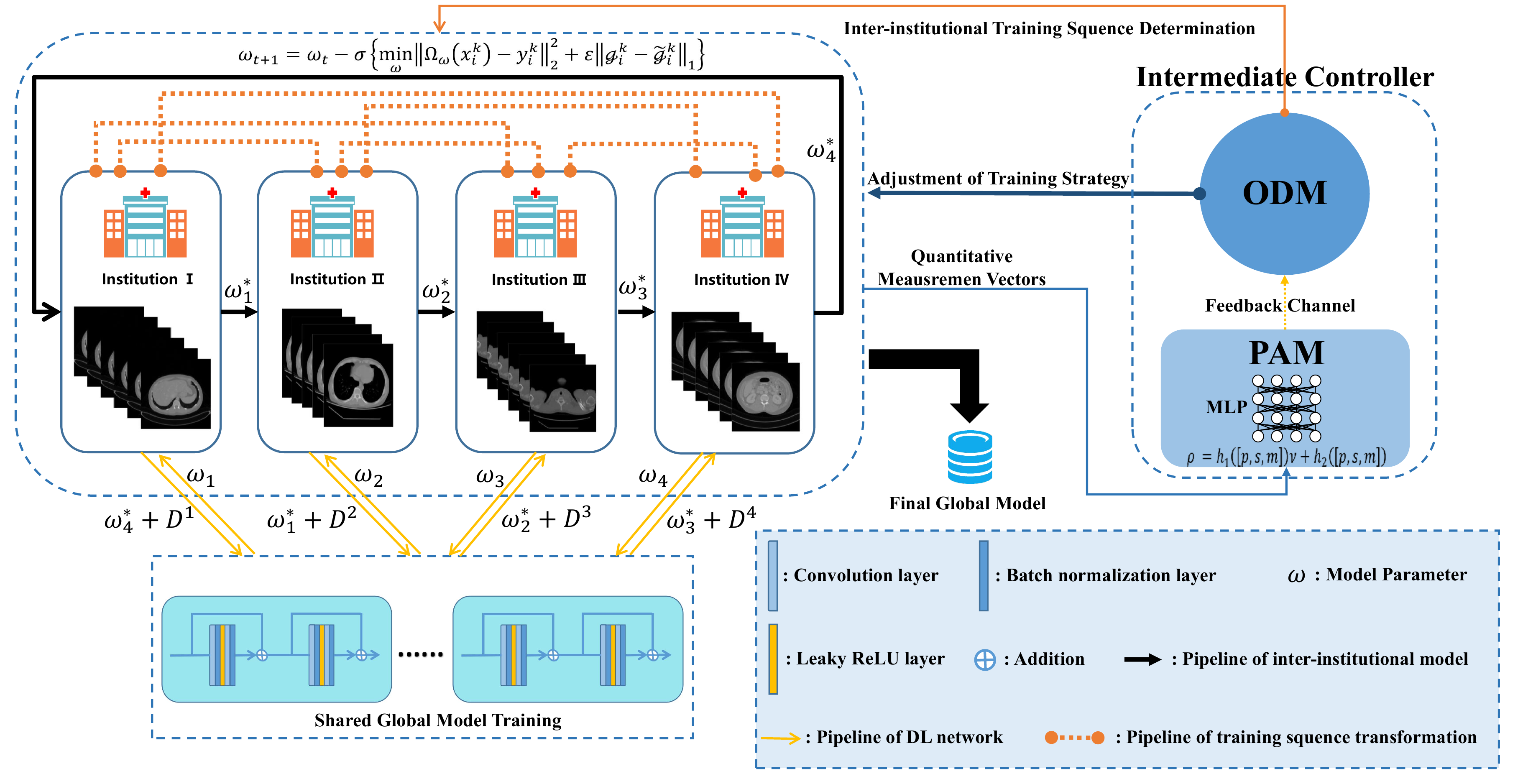}
\caption{The framework of the presented icP2P-FL.} \label{Framework}
\end{figure}

\section{Background}
Low-dose computed tomography (CT) imaging has been widely used in clinics. However, lowering the dose level can result in degraded image quality \cite{5}. Deep learning (DL)-based methods have been developed to improve low-dose CT image quality, including data-driven DL methods \cite{1}\cite{2}, and model-driven deep unrolling methods \cite{3}\cite{4}. However, most DL-based methods for low-dose CT imaging usually require large amounts of diversity-rich data which can be labor-intensive to collect. In addition, it is difficult to collect and share CT data from multiple institutions efficiently due to patient privacy \cite{7}, Moreover, a single-institution training network may suffer from inadvertent bias that ultimately limits imaging performance even on the data collected at the same institution due to the considerable heterogeneity exists among multiple institutions.

Federated learning (FL) enables decentralized model training across geographically dispersed multi-institutional data silos to address these limitations. In the FL,  distributed institutions collaboratively learn a shared model while keeping data local for privacy \cite{8}. The FL is also used in the medical imaging field and obtains better benefits over traditional DL-based methods \cite{11}\cite{12}\cite{13}\cite{14}. In centralized (center-to-peer) federated learning, a central server is required to orchestrate the process of training a single shared model, which needs high communication costs between the central server and multiple institutions. Various strategies have been investigated to reduce communication costs \cite{9}\cite{10}. Among them, the decentralized FL (peer-to-peer) model avoids the need for a central server to orchestrate the process and is more efficient for different issues specific to centralized FL models \cite{16}\cite{17}. In the peer-to-peer (P2P) FL model, to avoid the catastrophic forgetting problem, continual learning plays a critical role \cite{18}\cite{19}\cite{20}. The continual learning strategy in the P2P FL can train the model with the ability to remember old knowledge and learn new knowledge.

In this work, inspired by the P2P FL and continual learning strategy, we propose a novel P2P federated continual learning strategy to improve low-dose CT imaging performance from multiple institutions. The newly proposed method is termed as P2P continual FL with intermediate controllers, i.e., icP2P-FL. To the best of our knowledge, this is the first one to utilize P2P continual FL in low-dose CT image reconstruction. Specifically, the proposed icP2P-FL does not require a central server and adopts a P2P model where the institutions only communicate with their one-hop neighbors to reduce certain communication costs. In the proposed icP2P-FL method, due to the stable and effective common features among different institutions, all the institutions collaboratively share a global model and directly interact with each other without depending on a central server. Each institution can initiate an update process dynamically. Furthermore, an intermediate controller is developed to make the overall training more flexible. Due to the high frequency of interaction and the intermediate controller, the proposed icP2P-FL converges quickly. The CT dataset with different protocols from multiple institutions is collected to validate and evaluate the denoising performance of the proposed icP2P-FL for image reconstruction. Experimental results show that the proposed icP2P-FL method obtains competitive denoising performance in the CT reconstruction task compared with the other competing methods, and reaches an accuracy similar to a model trained with centralized data from all the institutions.

\section{Methods}
Let us consider an environment with $K$ institutions, where each institution has training data $D^{k} = \left\{ {x_{i}^{k},y_{i}^{k}} \right\}_{i = 1}^{n^{k}}$ and characteristic data $C^{k} = \left\{ {z_{i}^{k}} \right\}_{i = 1}^{n^{k}}$ with $n^{k}$ labeled samples.
\subsection{The description of icP2P-FL}
Figure 1 shows the framework of the presented icP2P-FL method. The presented icP2P-FL method consists of a peer-to-peer FL framework, an intermediate controller, and a DL network. There is no central server for coordinating the training process, and all the institutions are connected directly in a P2P manner. The training process of one cycle is conducted as the following steps:
\begin{enumerate}
\item The initial institution order is set (Initial\underline{\space}sequence: [Institution \uppercase\expandafter{\romannumeral1}, Institution \uppercase\expandafter{\romannumeral2}, Institution \uppercase\expandafter{\romannumeral3}, \dots, Institution k]), and network parameters are initialized.
\item The dataset $D^{1}$ of Institution \uppercase\expandafter{\romannumeral1} is used to train the shared global model $\Omega_{\omega}$ for CT image denoising.
\item After the training procedure is completed in Institution \uppercase\expandafter{\romannumeral1}, the trained model $\Omega$ is employed to evaluate the performance on the characteristic data from Institution \uppercase\expandafter{\romannumeral1} to obtain quantitative measurement vectors, i.e., $\left\lbrack {p,s,m} \right\rbrack_{i}^{k}$.
\item The parameters $\omega^{*}_{1}$ in the model are transferred to the one-hop neighbor, i.e., Institution \uppercase\expandafter{\romannumeral2}. $D^{2}$ is used to fine-tune the shared global model with the gradient correcting constraints $\left\| {g_i^k-{\widetilde{g}}_i^k} \right\|_{1}$ included to retain the significant model weights and archive the catastrophic forgetting compensation.
\item After the training is completed in Institution \uppercase\expandafter{\romannumeral2}, the trained model $\Omega$ is also employed to evaluate the performance on the characteristic data from Institution \uppercase\expandafter{\romannumeral2} to obtain quantitative measurement vectors, and the another institutions are trained similarly.
\item When all the institutions finish training, the intermediate controller grades all the quantitative measurement vectors $\left\lbrack {p,s,m} \right\rbrack$, of all the institutions on-line. This will immediately determine the inter-institutional training strategy and whether the global shared model is well-trained.
\end{enumerate}

\begin{algorithm}[htb] 
\caption{\quad icP2P-FL Training.} 
\label{alg:Framwork} 
\begin{algorithmic}[]
\REQUIRE ~~\\ 
\STATE Initialize the global model $\Omega$ parameters;
\STATE Initialize $K$ institution sequence: [Institution \uppercase\expandafter{\romannumeral1}, Institution \uppercase\expandafter{\romannumeral2}, Institution \uppercase\expandafter{\romannumeral3}, \dots, Institution k];
\STATE Datasets: $D^{k} = \left\{ {x_{i}^{k},y_{i}^{k}} \right\}_{i = 1}^{n^{k}}$ and $C^{k} = \left\{ {z_{i}^{k}} \right\}_{i = 1}^{n^{k}}$;
\STATE ~~\\
\textbf{Training:}
\noindent 
\WHILE{Transmission $\leftarrow1$ to $T$}
\WHILE {k $\leftarrow1$ to $K$}
\STATE Institution k training: ResNet $\leftarrow D^{k}$, and initial $\omega$
\WHILE {site-round $\leftarrow1$ to $S$}
\FOR {$i$ to $n^{k}$}
\STATE loss = $\left\| {\Omega_{\omega}\left( x_{i}^{k} \right) - y_{i}^{k}} \right\|_{2}^{2}$ $\leftarrow$ $x_{i}^{k}$ and $y_{i}^{k}$
\STATE loss $\leftarrow$ $\varepsilon\left\| {g_i^k-{\widetilde{g}}_i^k} \right\|_{1} = \left\| {\frac{\partial{\parallel \Omega_{\omega}(x_{i}^{k}) - y_{i}^{k} \parallel}_{2}^{2}}{\partial N_{y_{i}^{k}}} - QP(g_{i}^{k},G_{i}^{k})} \right\|_{1}$ 
\STATE $\omega_{t + 1} = \omega_{t} - \sigma\left\{ {\min\limits_{\omega}{\left\| {\Omega_{\omega}\left( x_{i}^{k} \right) - y_{i}^{k}} \right\|_{2}^{2} + \varepsilon\left\| {g_i^k-{\widetilde{g}}_i^k} \right\|_{1}}} \right\}_{t}$
\STATE Calculate: $\left\lbrack {p,s,m} \right\rbrack_{i}^{k}$ $\leftarrow$ the characteristic data $z_{i}^{k}$
\ENDFOR
\STATE Calculate: ${\sum_{i = 0}^{n}\left\lbrack {p,s,m} \right\rbrack_{i}^{k}}$
\ENDWHILE
\STATE ${k:}\left\lbrack {p,s,m} \right\rbrack$ $\rightarrow$ Intermediate controller $\stackrel{MLP}{\Longrightarrow}$ $\rho~^{k}$
\STATE $\omega_{k}^{*}$: Institution k $\rightarrow$ Institution k+1
\ENDWHILE
\IF{ $\rho~^{k}$ $\geq$ the set determination threshold}
\STATE Adjust institution sequence, the number of trans\_round and site\_round
\ENDIF
\ENDWHILE

\ENSURE ~~\\ 
\STATE The final shared global model $\Omega_{\omega}$ for each institution 
\end{algorithmic}
\end{algorithm}

It should be noted that the scope of the intermediate controller can help P2P-FL train the shared global model more flexibly and efficiently with the on-line assessment. After completing multiple cycles, we can obtain the well-trained share global model in the proposed icP2P-FL method to obtain high-quality low-dose CT images from all the institution efficiently.

\subsection{Inter-institutional Incremental Learning}
In this work, we adopt a cyclic task-incremental continual learning (CICL) to guarantee the model’s performance across multiple institutions in the transmission cycles. In incremental learning (ICL), multiple institutions continuously train a shared model in a set order. Each institution trains the shared model only once in a cycle and finally aggregates it to obtain the final model. Cyclic Incremental continual Learning (CICL) repeats the ICL process, i.e., cyclic training across multiple institutions and reducing forgetting by determining the number of rounds in each institution.

\begin{sloppypar}
Borrowing from the CICL strategy \cite{22}\cite{22}, the proposed icP2P-FL method adopts the regularization-based CICL to keep old knowledge in the previous institution and learn new knowledge in a new institution, and it can be expressed as follows:
\end{sloppypar}

\begin{equation}
    \min\limits_{\omega}{\left\| {\Omega_{\omega}\left( x_{i}^{k} \right) - y_{i}^{k}} \right\|_{2}^{2} + \varepsilon\left\| {g_i^k-{\widetilde{g}}_i^k} \right\|_{1}}
\end{equation}
\noindent where $x_{i}^{k}$ is the $i$th low-dose CT image, $y_{i}^{k}$ is the normal-dose CT images data in each institution dataset $D^{k} = \left\{ {x_{i}^{k},y_{i}^{k}} \right\}_{i = 1}^{n^{k}}$ in a batch, respectively.
$n^{k}$ is the number of samples in the $k$th institution. $\Omega_{\omega}$ is network with parameters $\omega$, and
$\omega_{k - 1}^{*}$ is the transmitted parameter of the previous institution after completing all training of the site-round.
$\left\| \cdot \right\|_{2}^{2}$ is the mean squared error (MSE) loss function, i.e., $\mathcal{L}_{2}$ norm. And $\varepsilon$ is the hyper-parameter,
$\left\| \cdot \right\|_{1}$ is the $\mathcal{L}_{1}$ norm and can be concretely expressed as follows:
\begin{equation}
\left\| {g_i^k-{\widetilde{g}}_i^k} \right\|_{1} = \left\| {\frac{\partial{\parallel \Omega_{\omega}(x_{i}^{k}) - y_{i}^{k} \parallel}_{2}^{2}}{\partial N_{y_{i}^{k}}} - QP(g_{i}^{k},G_{i}^{k})} \right\|_{1}
\end{equation}
where $g_i^k$ is the gradient measurements with respect to the $y_{i}^{k}$-th neuron $N_{y_{i}^{k}}$ of the last output layer in $\Omega_{\omega}$. $QP(·)$ is the quadratic programming method to correct the inter-institutional update directions, and $G_{i}^{k} = \lbrack g_{i}^{k - 1},g_{i}^{k},\omega - \omega_{k - 1}^{\ast}\rbrack$.

In the $t$th round, model parameters at the $k$th institution can be updated as follows:
\begin{equation}
    \omega_{t + 1} = \omega_{t} - \sigma\left\{ {\min\limits_{\omega}{\left\| {\Omega_{\omega}\left( x_{i}^{k} \right) - y_{i}^{k}} \right\|_{2}^{2} + \varepsilon\left\| {g_i^k-{\widetilde{g}}_i^k} \right\|_{1}}} \right\}_{t}
\end{equation}
where $\sigma$ is the learning rate at each institution.

\subsection{Intermediate controller}
The intermediate controller contains the performance assessment module and on-line determination module.
\subsubsection{Performance Assessment Module}
\begin{sloppypar}
The performance assessment module (PAM) allows for evaluating the denoising performance for image reconstruction with quantitative measures, i.e., peak signal-to-noise ratio (PSNR), structural similarity index measure (SSIM), and mean square error (MSE). To incorporate the three measurements into the proposed icP2P-FL method, the widely used  multilayer perceptron (MLP) is introduced as follows: 
\end{sloppypar}
\begin{equation}
    \rho~ = h_{1}\left( \left\lbrack {p,s,m} \right\rbrack \right)\nu + h_{2}\left( \left\lbrack {p,s,m} \right\rbrack \right),
\end{equation}
where $\nu$ is the image feature map of the network layer, $h_{1}$, $h_{2}$ are MLPs with shared parameters, and p, s, m are the metrics of PSNR, SSIM, and MSE, $\rho$ is the score of the entire dataset of each institution. The performance assessment module of the proposed icP2P-FL is pre-trained on labeled image data from each institutions, which consists of four fully connected layers.

\subsubsection{On-line determination module}
As the number and distribution of training datasets in each institution varies a lot, there are real differences in the computational efficiency and catastrophic forgetting rate of training the shared global model on different datasets. Then we developed the Online Decision Module (ODM), which relies on the scores obtained from the PAM by evaluating the model on each institution's characteristic dataset. The ODM enables degrade the all the quantitative measurement vectors $\left\lbrack {p,s,m} \right\rbrack$, of all the institutions on-line, and then immediately determines inter-institutional training order. The ODM can adjust the number of transmissions between two adjacent institutions (i.e., the number of cycles in the training stage). And it also determines the site-round at each institution (i.e., the number of training rounds in one cycle). Finally, it can finally determine whether the global shared model is well-trained.

\subsection{Implementation Details}
In this work, the backbone DL network of the presented icP2P-FL method adopts a modified residual network (ResNet) with 12 residual blocks and 2 residual layers \cite{22}. The training parameters are set as follows: (1) the number of transmissions and the number of site-round of each institution are respectively set to 10 and 5, (2) the round number is set to 100 and the weight is decayed at the 100th round by multiplying 0.2, (3) the learning rate and batch size of the model are $1.0\times10^{-4}$ and 64, respectively, (4) all the institutions have labeled data, i.e., normal-dose CT images/corresponding low-dose ones acquired with different protocols, (5) switch and fine-tuning mode are initially set to the Ture, (6) determination threshold is set to 1.4759, (7) the training image patches are set to $64\times64$ with a stride of 64. All the networks in this work are implemented with Pytorch library, and the FBP algorithm is based on the ASTRA toolbox by utilizing one NVIDIA Tesla V100s graphics processing unit (GPU) which has 32 GB memory capacity.
\begin{sloppypar}
    The presented icP2P-FL method is compared to several widely used denoising algorithms for image reconstruction, including the FBP algorithm, FedAvg \cite{15}, CL-SI (i.e., single institution model with centralized learning trained on institution-specific dataset), CL-MI (i.e., multiple institutions model with centralized learning trained on all datasets), and Semi-centralized Federated Learning network (SC-FL) \cite{24}. Three measures, i.e., PSNR, SSIM, and MSE, are employed to quantitatively evaluate the performance of the proposed our proposed icP2P-FL method and the comparison models.
\end{sloppypar}

\section{Experimental Results}
\subsection{Dataset}
To validate and evaluate the denoising performance of the proposed icP2P-FL method for image reconstruction, four different CT datasets are used in the experiments. Specifically, with the approval of the medical ethics committee of the local institution, three CT datasets were collected from three local hospitals, i.e., Institution \uppercase\expandafter{\romannumeral1}, Institution \uppercase\expandafter{\romannumeral2}, and Institution \uppercase\expandafter{\romannumeral3}. The three datasets are acquired with different protocols with different scanners. The fourth CT dataset is from the "2016 NIH-AAPM-Mayo Clinic Low Dose CT Grand Challenge" \cite{21}, termed as Institution \uppercase\expandafter{\romannumeral4}. All the CT images are served as normal-dose ones in this study. Then we simulated low-dose CT images (i.e., quarter dose) from the corresponding CT images based on the previous study \cite{23}. 
In Institutions \uppercase\expandafter{\romannumeral1}, \uppercase\expandafter{\romannumeral2}, and \uppercase\expandafter{\romannumeral3}, 800 and 100 cases are assigned for training and testing datasets, respectively. 200 and 100 cases are assigned for training and testing in Institution \uppercase\expandafter{\romannumeral4}.
To select characteristic datasets for validation evaluation, we employ 100 cases of paired images from each institution dataset independent of training and testing, with deeper features and individualized data distributions in each institution.
\renewcommand{\floatpagefraction}{.9}
\begin{figure}
\includegraphics[width=1.0\textwidth,height=0.6\textwidth]{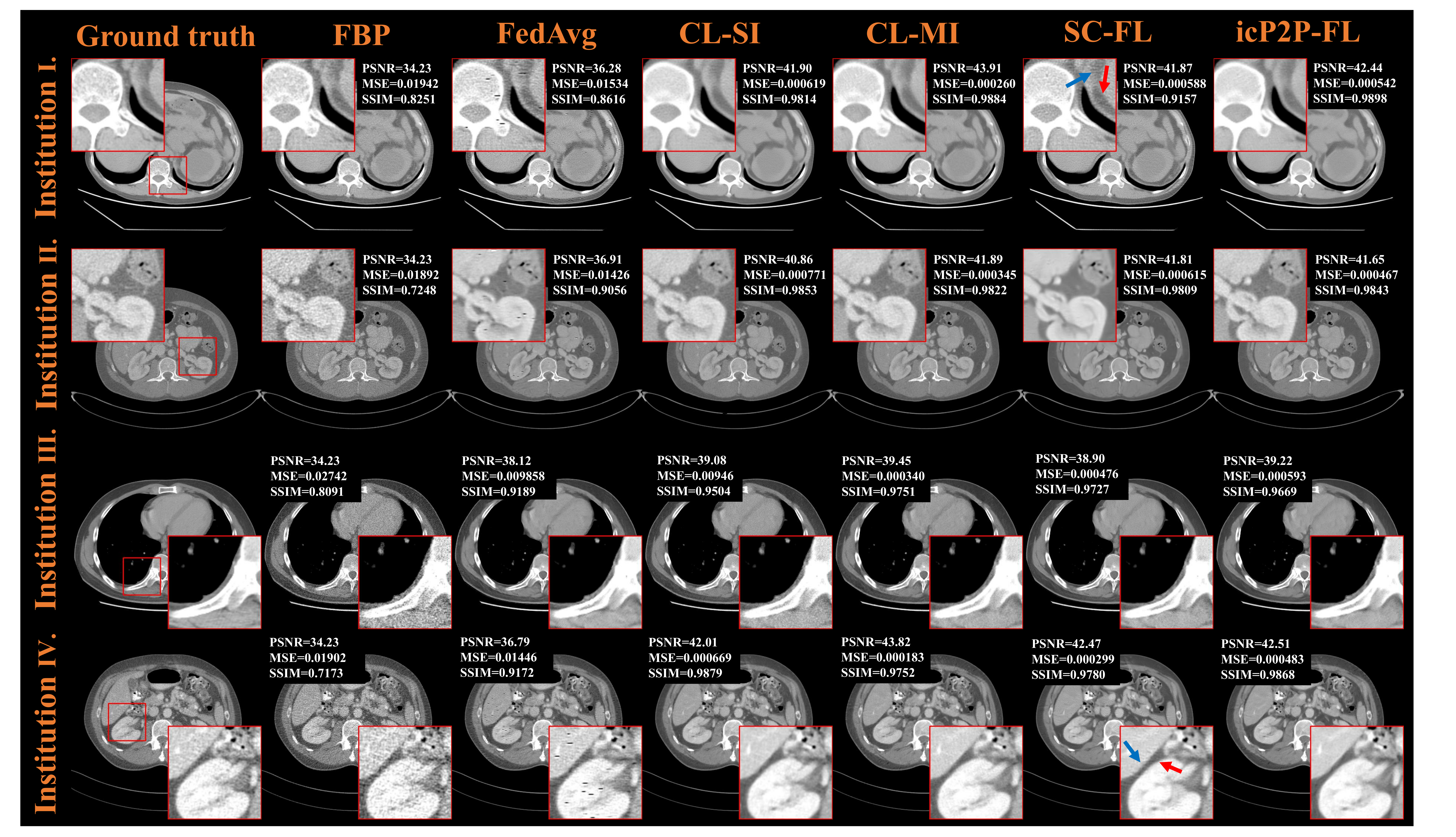}
\caption{Qualitative comparison of the image reconstructed by the different methods. The display windows for CT images at Institution 1, Institution 2, Institution 3 and Institution 4 are [-50, 50], [-100, 100], [-50, 50], [-150, 150] HU, respectively. The display windows for zoomed-in ROIs are [-10, 50], [-20, 80], [-30, 50], [-80, 100] HU, respectively.} \label{visualization-p2p_1}
\end{figure}

\subsection{Performance Comparison}
Figure 2 shows the example slice denoising with each individual method at four institutions and a corresponding zoomed-in region-of-interest (ROI) view indicated by the red boxes. The zoomed-in ROI view is used shed more light on the denoising performance for image reconstruction. The normal-dose CT images are served as ground truth for comparison. It can be seen that the low-dose FBP images contain severe noise-induced artifacts. FedAvg can reduce noise-induced artifacts to some extent, but it introduces undesired artifacts in the final results. The main reason is that considerable heterogeneity exists among the four institutions, and FedAvg fails to fully take the considerable heterogeneity into consideration. The CL-SI is trained on the institution-specific CT data and it obtains promising results in noise-induced artifacts suppression, but it can not process low-dose CT images from other institutions shown in Fig. 3. The degraded denoising performance for image reconstruction is attributed to the heterogeneity between the training dataset from one specific institution and the testing data from the other institutions. The CL-MI is trained on the pooling CT data from the four institutions, and it achieves the best performance in noise-induced artifacts removal and structure details recovery due to it considers all the latent characteristics in the four institutions. The SC-FL has better performance than the FedAvg and CL-SI by constructing high-quality labels in the central server. But it can be seen that the SC-FL results still contain residual noise-induced artifacts as indicated by the red and blue arrows due to the L1 loss function in the SC-FL.

The proposed icP2P-FL method provides a better denoising effect of image reconstruction, and obtains similar denoising and generalization performance visually with the CL-MI method, as shown in the zoomed-in ROIs. The main reason could be that the continuous learning strategy with an intermediate controller for training tuning in the proposed icP2P-FL allows the global model to perform multiple denoising tasks for image reconstruction, i.e., the CT data acquired with different protocols from different protocols scanners, and reduces the side effects of knowledge forgetting. Moreover, the quantitative assessments indicate that the proposed icP2P-FL method obtains superior performance which is similar to the CL-MI method, highlighting that the proposed icP2P-CL reaches an accuracy similar with the CL-MI method trained with pooling data from all the institutions.

\section{Conclusion}
This work proposes icP2P-FL as a novel P2P federated learning framework for training global models for CT image denoising and reconstruction from different institutions simultaneously. The motivation behind the proposed icP2P-FL is to enable efficient and effective denoising of CT images with large heterogeneity directly from different protocols and different scanners. The experimental results show that the proposed icP2P-FL method can achieve better performance for robust cross-institution CT denoising for image reconstruction than the existing methods, and reaches an accuracy similar to a model trained with pooling data from all the institutions. In future studies, more real clinical data should be enrolled to further demonstrate its performance.
\renewcommand{\floatpagefraction}{.9}
\begin{figure}
    \centering
    \includegraphics[width=0.8\textwidth,height=0.30\textwidth]{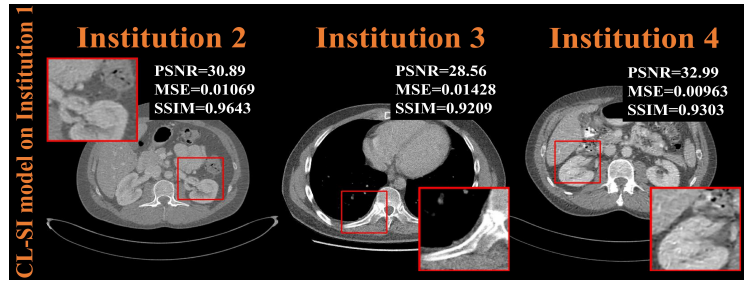}
    \caption{Results of institution 2, 3, 4 tested by the CL-SI model on Institution 1} \label{visualization-p2p_2}
    \end{figure}

\begin{table}[t]
    \caption{Mean and standard for PSNR and SSIM of institutions.}
    \label{psnr and ssim}
    \centering
    \begin{minipage}[t]{0.8\linewidth}
        \resizebox{\textwidth}{!}{
            \begin{tabular}{ccccc}
            \hline
            \textbf{PSNR} & Institution 1 &\quad Institution 2 &\quad Institution 3 &\quad Institution 4\\
            \hline
            FBP & 34.12$\pm$1.2814 &\quad 34.43$\pm$1.3234 &\quad 33.98$\pm$1.3848 &\quad 34.93$\pm$1.3176 \\ 
            FedAvg & 35.11$\pm$1.2679 &\quad 36.39$\pm$1.0313 &\quad 36.95$\pm$1.4068 &\quad 36.79$\pm$1.0041 \\
            Single\underline{\space}institution & 41.01$\pm$0.2628 &\quad 39.96$\pm$0.2920 &\quad 39.02$\pm$0.2182 &\quad 41.75$\pm$0.3012 \\
            Multi\underline{\space}institutions & 43.17$\pm$0.2619 &\quad 41.83$\pm$0.3237 &\quad 39.33$\pm$0.4146 &\quad 42.91$\pm$0.2751 \\
            SC-FL & 41.75$\pm$0.2679 & \quad42.09$\pm$0.3313 &\quad 38.09$\pm$0.4068 &\quad 42.64$\pm$0.3041 \\
            \textbf{icP2P-FL} & 41.78$\pm$0.2628 &\quad 42.55$\pm$0.2920 &\quad 39.13$\pm$0.2182 &\quad 42.86$\pm$0.3012 \\
            \hline
        \end{tabular} }
    \end{minipage}
\hfill
    \begin{minipage}[t]{0.8\linewidth}
        \resizebox{\textwidth}{!}{
            \begin{tabular}{ccccc}
            \hline
            \textbf{SSIM} & Institution 1 &\quad Institution 2 &\quad Institution 3 &\quad Institution 4\\
            \hline
            FBP & 0.8316$\pm$0.03167 &\quad 0.7563$\pm$0.03559 &\quad 0.7936$\pm$0.02883 &\quad 0.7152$\pm$0.02605 \\ 
            FedAvg & 0.8641$\pm$0.02679 &\quad 0.8933$\pm$0.02254 &\quad 0.9017$\pm$0.02387 &\quad 0.9063$\pm$0.01998 \\
            Single\underline{\space}institution & 0.9864$\pm$0.00763 &\quad 0.9837$\pm$0.00463 &\quad 0.9897$\pm$0.00627 &\quad 0.9752$\pm$0.00798 \\
            Multi\underline{\space}institutions & 0.9867$\pm$0.00323 &\quad 0.9826$\pm$0.00485 &\quad 0.9763$\pm$0.00935 &\quad 0.9747$\pm$0.00845 \\
            SC-FL & 0.9726$\pm$0.00988 & \quad0.9779$\pm$0.00818 &\quad 0.9766$\pm$0.00900 &\quad 0.9722$\pm$0.00960 \\
            \textbf{icP2P-FL} & 0.9833$\pm$0.00881 &\quad 0.9790$\pm$0.00761 &\quad 0.9737$\pm$0.00852 &\quad 0.9781$\pm$0.00886 \\
            \hline
        \end{tabular} }
    \end{minipage}
\end{table}

\bibliographystyle{splncs04}
\bibliography{references.bib} 

\end{document}